# Prophet: A Speculative Multi-threading Execution Model with Architectural Support Based on CMP

Dong Zhaoyu, Gao Bing, Zhao Yinliang, Song Shaolong, Du Yanning
(Department of Computer Science and Technology, Xi'an Jiaotong University, Xi'an 710049, China)

**Abstract**-- Speculative Multithreading (SpMT) has been proposed as a perspective method to exploit Chip Multiprocessors (CMP) hardware's potential. It is a thread level speculation (TLS) model mainly depending on software and hardware co-design. This paper researches speculative thread-level parallelism of general-purpose programs and a speculative multi-threading execution model called Prophet is presented. The architectural support for Prophet execution model is designed based on CMP. In Prophet the inter-thread data dependences are predicted by pre-computation slice (p-slice) to reduce RAW violation. Prophet Multi-versioning Cache system along with thread state control mechanism in architectural support are utilized for buffering the speculative data, and a snooping bus based cache coherence protocol is used to detect data dependence violation. The simulation-based evaluation shows that the Prophet system could achieve significant speedup for general-purpose programs.

**Key words: Speculative Multi-threading, Execution Model, Architectural Support**

## 1 Introduction

Exploiting program parallelism is one of the most effective ways to improve the program execution performance. Traditionally, microprocessors and compilers improve the execution performance primarily by exploiting instruction-level parallelism (ILP). However, the amount of ILP that can be extracted from a program is limited by instruction window size, clock cycle and memory latency [1, 2]. One promising technique for overcoming this problem is thread-level-parallelism (TLP) which use multiple processing elements on a single chip to execute multiple threads from the same program in parallel. The automatic parallelism has made great progress in numeric programs. But for general-purpose programs it can not exploit sufficient parallelism [7, 8]. Since the irregular control flow and data flow in general-purpose programs make it hard for compiler to find enough threads which can execute in parallel without violating dependences.

Speculative multi-threading (SpMT) exploits program's thread level parallelism and it is an effective approach to improve the execution performance of general-purpose programs. SpMT can be implemented exclusively in software [3] or hardware [6]. Prophet SpMT execution model proposed in this paper uses a co-designed scheme which makes use of hardware and software to improve performance of sequential program. The Prophet compiler is used to divide a sequential program into threads which can be executed parallel in the absence of ambiguous data and control dependences, that is, these threads are speculatively paralleled. Upon dependence violations, the underlying architectural support will detect it and recover from mis-speculation and resume execution again.

As to SpMT execution model, paper [3] proposed CorD model which is implemented completely in software. In this model, speculative thread will be discarded instead restarted once dependences occur. The data modified correctly in the thread will be copied into memory. This method somewhat increases runtime overhead and limits the thread size. The Pinot model presented in paper [4] prescribes that each thread can only spawn one sub thread in its life time. If a thread spawns its second sub thread then the first one will be squashed implicitly. This approach is helpful to reduce dynamic instructions but somewhat limits the utilization of multi-core hardware. In addition Pinot model transfers register value at runtime by register time. This method introduces the false RAW violation while decreasing register transfer overhead and makes the hardware more complex. Using pre-computation technology proposed in paper [5, 9], we design the Prophet model with architecture support based on CMP, where the program's correctness is guaranteed by the thread validation and commition mechanism. The thread

Supported by National High Technology Research and Development Program of China under Contract No.2008AA01Z136

commits its data to memory only when speculation successes and this is supported by hardware. Thus it is more effective than that implemented by software. Prophet get free of runtime register transfer overhead by pre-computation, and register false RAW violation is eliminated meantime.

Most SpMT system relies on compiler to generate parallel threads [4, 5, 10, 11]. We develop Prophet compiler to divided the sequential program into parallel threads despite uncertainty whether these thread are actually independent or not.

The remainder of the thesis is organized as follows. Section 2 describes Prophet execution model. Section 3 is a brief description of Prophet compiler. In Section 4 we describe Prophet architecture. We evaluate our approach in section 5 and summarize the paper in section 6.

## 2 Prophet Execution Model

The execution model of Prophet is targeted by Prophet Compiler and implemented by the underlying Prophet Architecture support. It provides a mechanism to maintain program order and defines the semantic of thread spawn, squash and restart operations so that threads can be transferred to the right state and correctness of the executed program can be maintained. In Prophet, every thread can spawn any number of sub threads. All the sub threads are spawned in an out-of-order manner. Threads commit data in their sequential order. When a thread is guaranteed not to violate any data dependences with its logically-earlier threads, it can commit all its speculative modifications to the main memory. In the case when the speculation fails, all the logically-later threads currently running are all squashed and the failed thread itself is restarted.

### 2.1 Parallel Execution in Prophet

Sequential programs are partitioned by Prophet compiler and each thread in Prophet executes one portion. A thread is identified with a Spawn Point (SP) and a Control Quasi-independent Point (CQIP). Each thread predicts its live-ins by executing a pre-computation slice (p-slice) which is statically generated by Prophet compiler. Fig.1 shows the parallel threads in Prophet comparing with their sequential execution. In Fig.1 thread 1 is the stable thread and others are speculative threads. There is only one stable thread in the system which is safe and will verify its immediate successor on meeting its CQIP. If the successor is verified to be right and have not violated any dependence, the stable thread will commit all its data and pass a stable token to the successor before it quits. Then the successor which received the stable token will become the new stable thread. Each thread will spawn a sub thread when it meets a SP. The speculative threads will stall on meeting their CQIP and wait for the verification by the stable thread. If the verification fails, all their result will be discarded and the stable thread will continue to run.

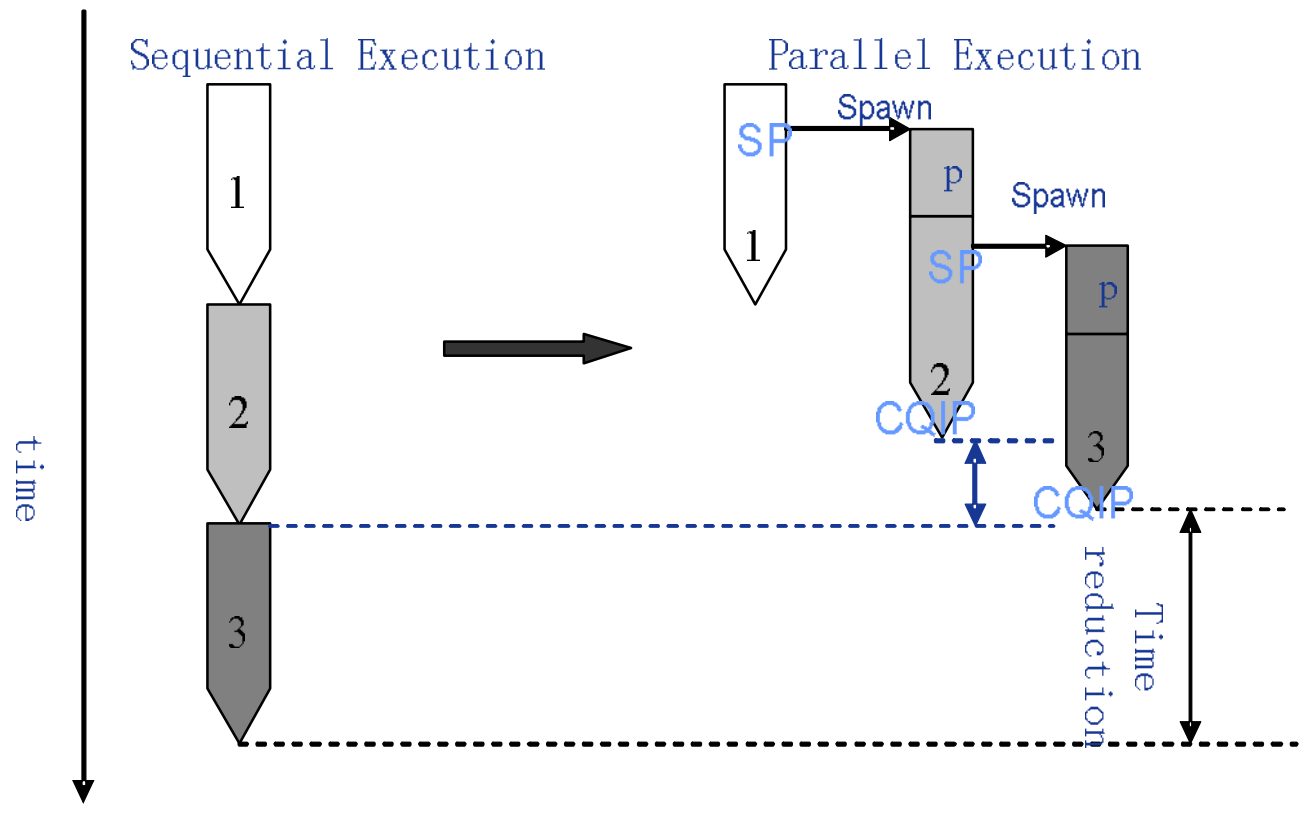

Fig.1 Parallel Execution in Prophet

### 2.2 Thread States in Prophet

Figure 2 illustrates the thread state transition in Prophet. The figure depicts all the possible state in a thread life cycle. Table 1 summarizes all the states and their semantics in execution model.

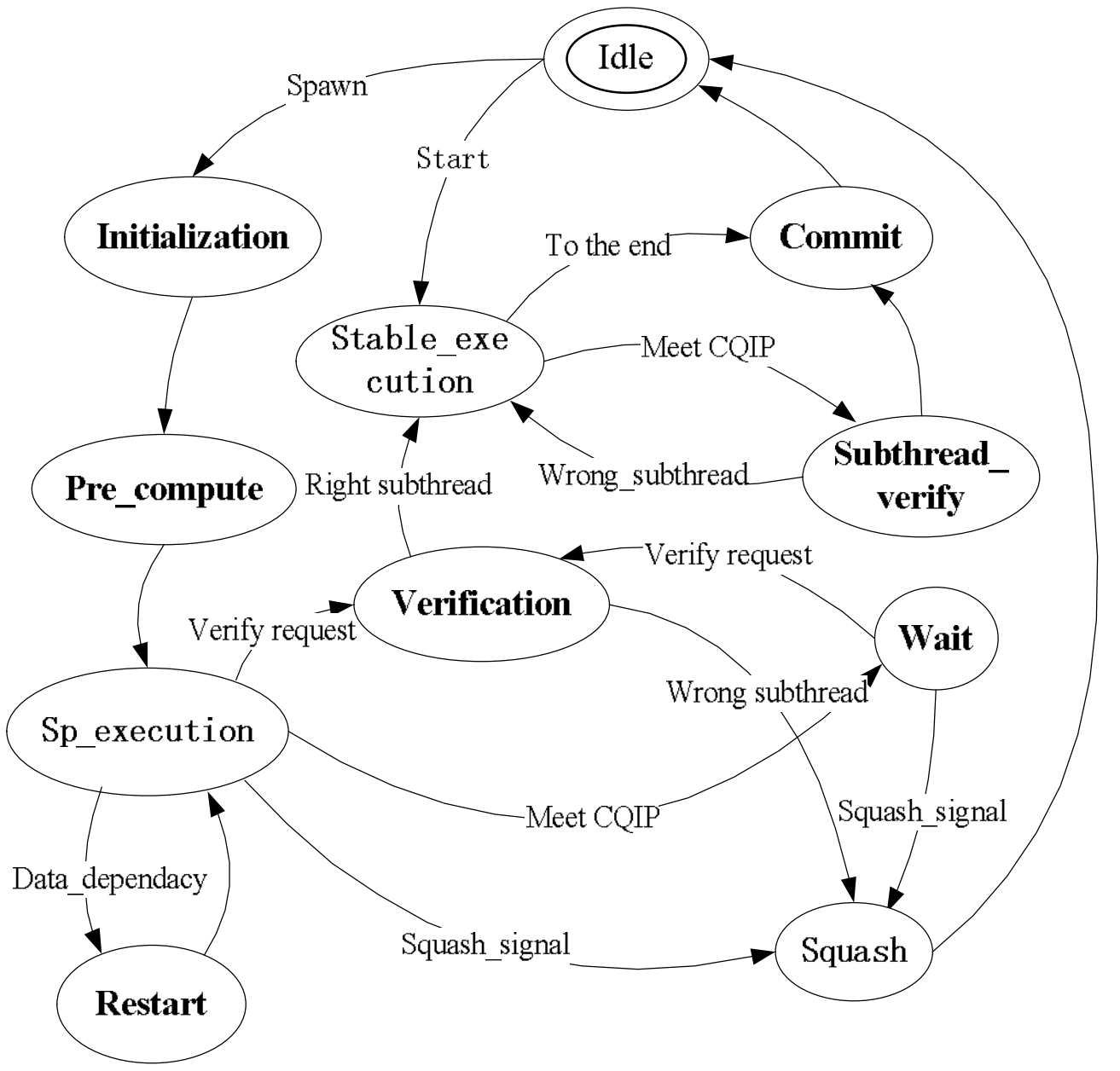

Fig.2 Thread State Transition Diagram

Table 1 Semantic of Thread States

| State | Semantic |
|---|---|
| Idle | The thread is free and has no task |
| Commit | The thread violates neither data nor control dependence and makes its speculative modifications visible to memory. After committing it sends a stable token to its immediate successor. |
| Sub thread verify | The parent thread executes a cqip instruction and send a verification message to its successive thread to verify the sub thread's pre-computation result |
| Wait | The thread has finished its execution and is blocked until it passes verification and receives a stable token |
| Squash | The thread fails to speculate and receives a squash message, all its sub-threads including itself are terminated, all the speculative results are discarded and all resources it possessed are freed |
| Verification | The thread receive a verification message and all its pre-computation results including memory data and register data are verified by its parent thread |
| Stable execution | The thread passes verification and receives a stable token and begins to execute in a stable manner, all its modifications needn't to be verified any more. |
| Initialization | The thread is newly spawned and begins to receive initialization to execute, including setting the start address, copying register values and thread version from its parent thread. |
| Pre-compute | The thread begins to execute the pre-computation slice. All the data generated in pre-computation state need verification before committing. |
| Sp_execution | The thread begins to speculatively execute. |
| Restart | The thread receives a violation message and squashes all its sub-threads and restores the register values to re-execute from its original start address. |

### 2.3 Interface

Table 2 Instruction Definition for Speculation

| instruction | operant | instruction description |
|---|---|---|
| spawn | *label* | Thread creation instruction; annotation specified the entrance *label* of new thread |
| cqip | *label* | Thread end instruction; annotation specified the entrance *label* of new thread |
| sqush | *label* | Thread squash instruction; annotation specified the *label* of squashed thread |
| pslice_entry | *label* | p-slice entry instruction; annotation specified the entrance *label* of new thread |
| pslice_exit | *label* | p-slice exit instruction; annotation specified the entrance *label* of new thread |

We have architected Prophet system to involve both the compiler and hardware, hence we require an interface between them. In Prophet the interface is defined at assembler language level and is supported directly by the underlying architecture. Table 2 lists all the instructions used in Prophet for thread speculation.

## 3 Compiler Support

The profile-guided Prophet compiler is developed to support the execution model. The main task of Prophet compiler is to parallelize the sequential program using the interface defined by Prophet execution model. So that the program can be executed in a speculatively parallel manner and the thread level parallelism is exploited. The compiler peels off threads speculative mainly from loop region and function calls. By analysis the Control Flow Graph (CFG), each basic block is considered as a candidate thread.

In addition, Prophet compiler also generates p-slice for each thread. A thread's p-slice can be viewed as a simplification of the code segment in its parent thread after its SP. Each thread executes p-slice to predict its live-ins thus data dependence violation can be reduced.

## 4. Prophet Underlying Architecture

The goal of Prophet architecture is to support Prophet execution model at runtime. The normal tasks performed by architecture to support speculation include: (1) speculative result buffering; (2) memory dependence violation detection and recovering; (3) register synchronization. As to Prophet that pre-computation is supported, its architecture must distinguish between the read/write operation in pre-computation phase and regular speculation phase. This section is organized as follow: first the Prophet architecture support is outlined at a higher level. Then the basic mechanism to support pre-computation is presented. Finally, the detailed design of multi-version cache is described including memory data L1 cache and register cache to explain how Prophet buffers speculation result by read and write operations.

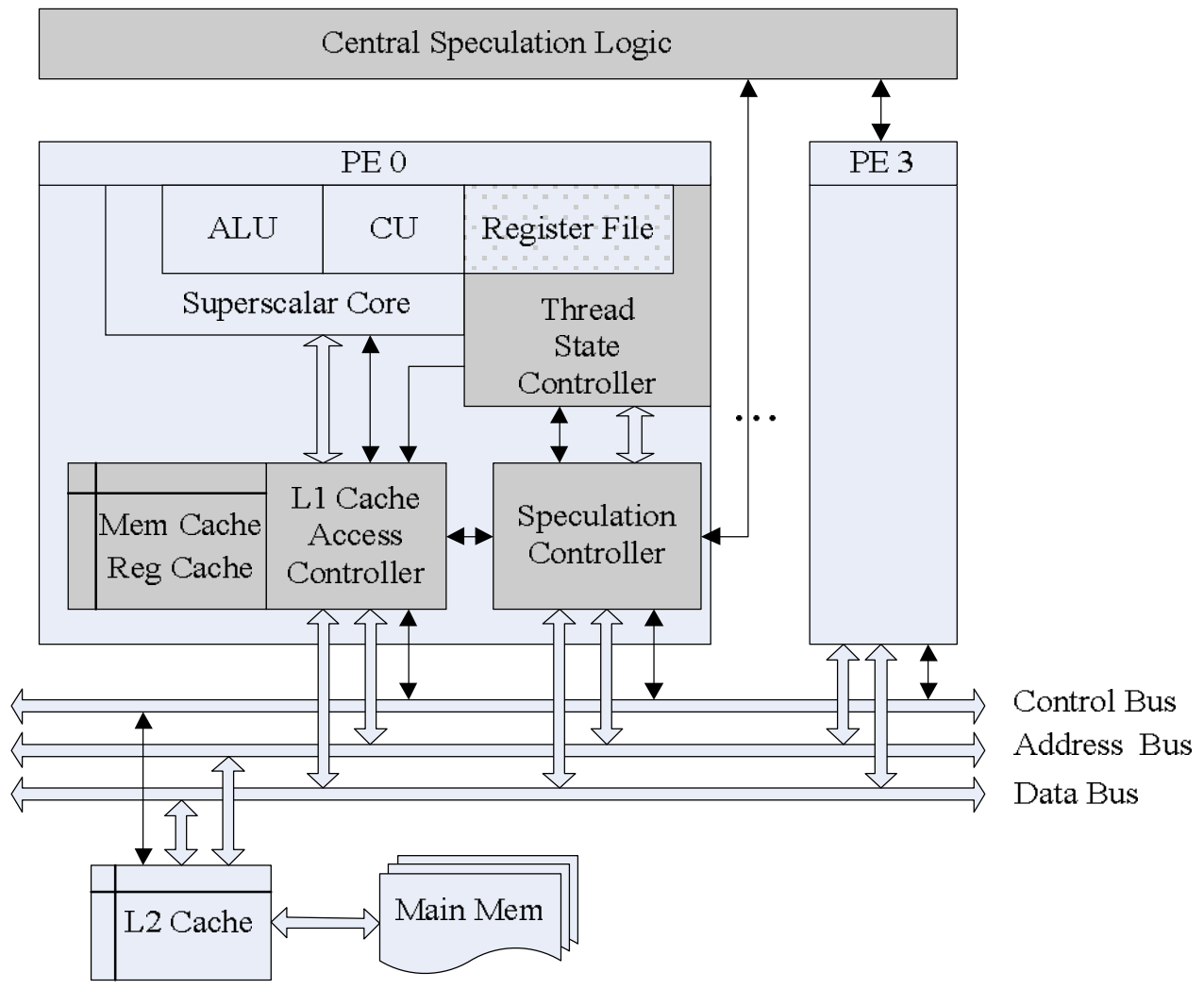

Fig. 3 Prophet Architecture

Prophet architectural support is loosely derived from Hydra [12] by adding the components to support speculative multi-threading execution. Fig.3 shows the anatomy of the Prophet Architectural support. The blocks appearing in a color shade of gray are components enabling speculation in Prophet, including L1 memory data and register cache, thread state controller and central speculation logic. Prophet buffers the speculation results, performs memory communication and disambiguation through L1 memory data cache. A single snoopy bus protocol is used to maintain L1 cache coherence. Thread state controller insures thread to transfer to the correct state when some events occur. Fig.4 shows the internals of a conceptual PE supporting speculation. The shadow blocks are extended in a normal PE. They are as follows: (1) instruction decode logic to decode the instructions defined in Prophet execution model, (2) register cache and L1 memory cache to maintain the speculation result, (3) thread controller.

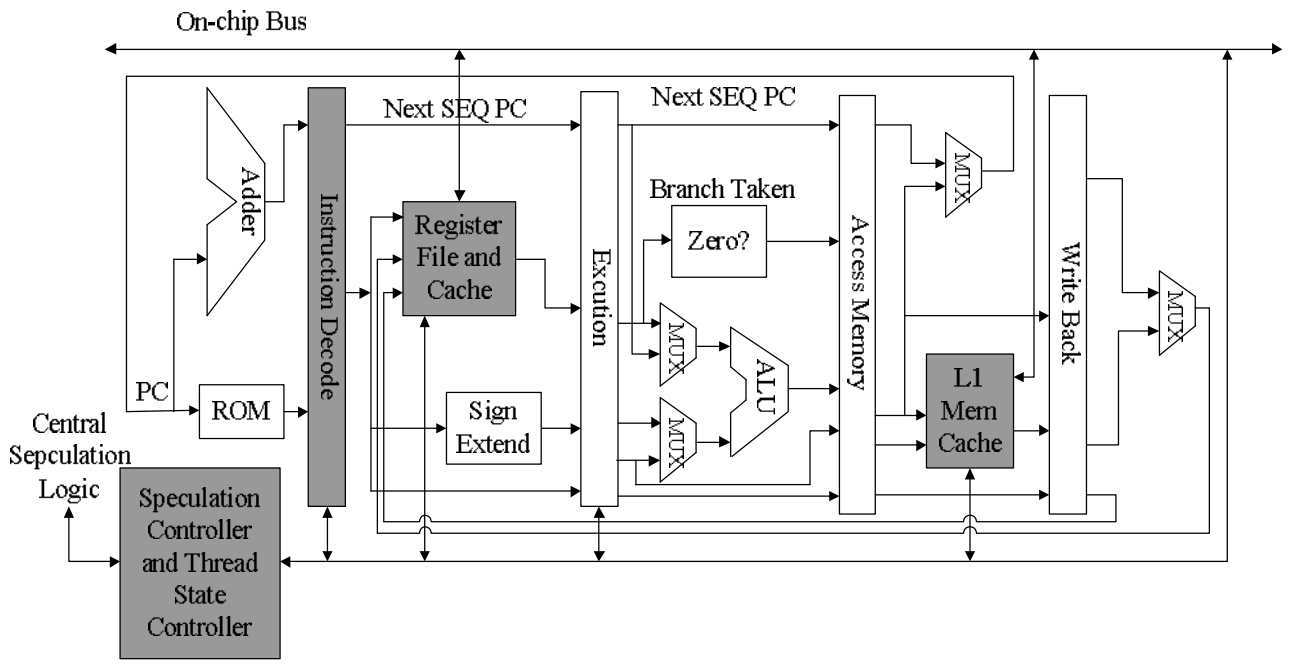

Fig.4 Block Diagram of the Inside of a Conceptual PE

### 4.1 Pre-computation Support

The most significant difference between pre-computation and speculation is that they have different view of the data generated by the parent thread. Fig.5 depicts such circumstance, thread0 writes const1 to X and than changes the value to const2 after spawning thread1. The spawned thread1 needs to read X both in pre-computation and speculation. The value read by thread1 in pre-computation should be const1 and that in speculation should be const2. Another difference is that they have different view of the logical-earlier threads. For a sub thread in pre-computation, the threads visible for it only include its parent and the parent's logical-earlier threads. But for a sub thread in speculation, all its logical-earlier threads are visible.

In order to support pre-computation in Prophet, we introduce thread version to distinguish values generated for the same variable in individual threads. In our scheme, thread version is a positive integer. When a thread is in pre-computation, all the data generated is marked with version zero. When the thread is in speculation state, all the data is marked with its thread version. On spawning, the parent thread passes its thread version to the sub thread and increases its own thread version by one. When the sub thread reads the data generated by its parent thread, it can find the right one by comparing data versions. The data the sub thread should read in pre-computation from its parent thread is the one with the max version less than or equal the sub thread's version.

In Prophet, every thread is spawned in an out-of-order manner. So it is convenient to use Immediate Successor List (ISL) [12] to maintain the sequential order. When a thread is spawned in an ISL, it will inherit all its parent's successors and become the immediate successor of its parent thread. The most speculative thread's successor is set to null. It is also very easy to use ISL to support pre-computation. In our scheme every sub thread keeps the hardware pointer to its parent thread. Thus, when a thread is in pre-computation, only its parent and all its parent's predecessors in ISL will be accessible. When a thread is in speculation, all its predecessors in ISL will be accessible.

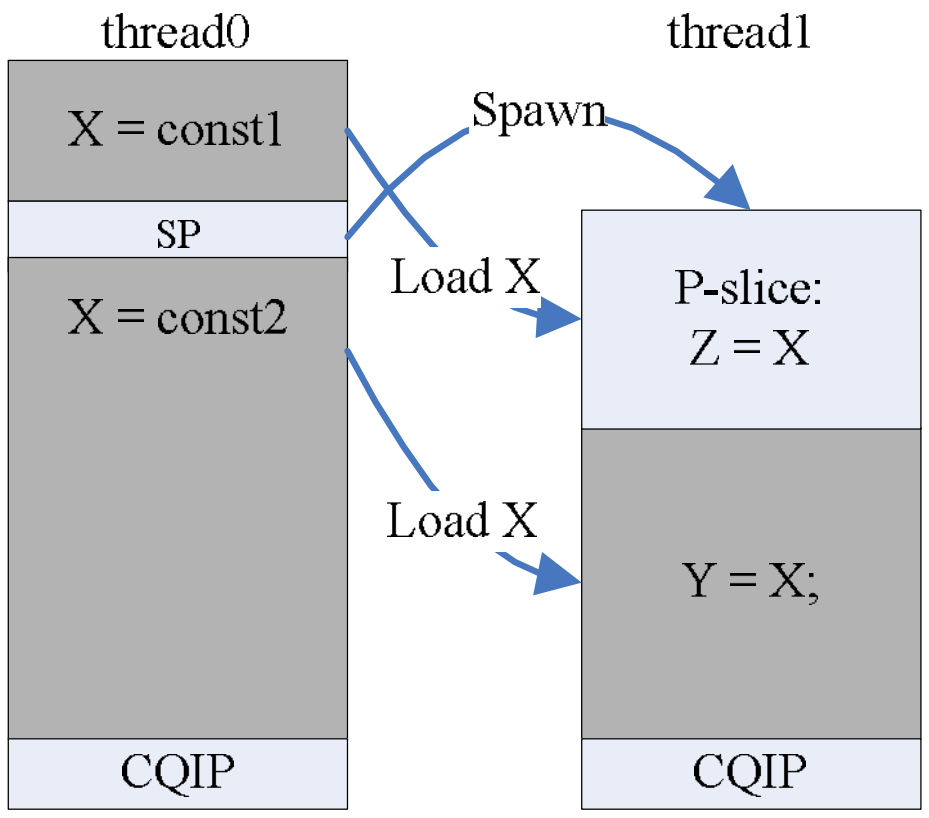

Fig.5 Different Data Are Visible in Pre-computation and Speculation

### 4.2 Speculative Result Buffering

Prophet employs cache system to buffer speculative result and resolve data dependency violation. The cache system consists of L1 memory data cache and register cache. The memory data cache is similar to Multi-Version Cache [13] and capitalizes on a consistence-based cache protocol to detect violation. Register cache actually is a multi-version register file. Since p-slice is constructed for each thread to reduce data dependence violation, Prophet must track multi-versioned data not only among different threads but also within each individual thread. Prophet performs memory read and write operations in L1 cache according to speculative coherence scheme. For speculation to succeed, any cache line with the newest version is committed to the memory when the thread transfers to stable state. If speculation failed, the thread is squashed and all the cache lines are invalided en masse.

#### 4.2.1 Memory Data L1 Cache Line

In Prophet, a memory cache line can be in one of such states: invalid, exclusive (Ex), share (Sh) or old (O). The invalid state indicates that the content in the cache line is no longer valid and should not be used any more. The exclusive state indicates that the cache line is originally generated by current thread. The share state denotes that this cache line is originally read from its predecessors. While the old state indicates that this cache line is old and there exists its new version. According Prophet's thread state, we extend the four basic cache line states as shown in Table 3. Memory cache needs to track different version of the same variable. When a write operation is performed in current thread, memory cache must check whether a new version will be generated. If so, the former version will be set obsolete and a new version is created.

Table 3 States of Memory Cache Line

| state | description |
|---|---|
| PreSh | read from predecessor in pre-computation |
| PreEx | generated by current thread in pre-computation |
| PreExO | generated by current thread in pre-computation and its version is old |
| SpSh | read from predecessor in speculation |
| SpShM | read from predecessor in speculation and modified |
| SpShO | read from predecessor in speculation and its version is old |
| SpEx | generated by current thread in speculation |
| SpExO | generated by current thread in speculation and its version is old |
| Invalid | cache line is invalid |

Table 4 Cache Message

| message | parameter | sender | sender state | operation |
|---|---|---|---|---|
| LPrR | address | current thread | pre-computation | read |
| RPrR | speculative level, address, thread version | successor | pre-computation | read |
| LPrW | address, value | current thread | pre-computation | write |
| LSpR | address | current thread | speculation | read |
| LSpW | address, value | current thread | speculation | write |
| RSpR | speculative level, address | successor | speculation | read |
| VioTest | speculative level, address | predecessor | stable or speculation | violation detection |

The basic message memory cache received can be write, read or violation detection. These messages may come from the current thread or its predecessors or successors. Table 4 enumerates all these messages and their parameters. Our memory cache accessing mechanism is summarized in a state transition diagram shown in Fig.6. When writing a new version of a cache line, the former cache line will be set old and a new cache line with new version will be created. A thread in speculation will send a VioTest message every time it writes a cache line. When performing a read operation, if the read hits then the cache line with newest version will be read, otherwise the reader will send a RSpR message if in speculation or a RPrR message if in pre-computation.

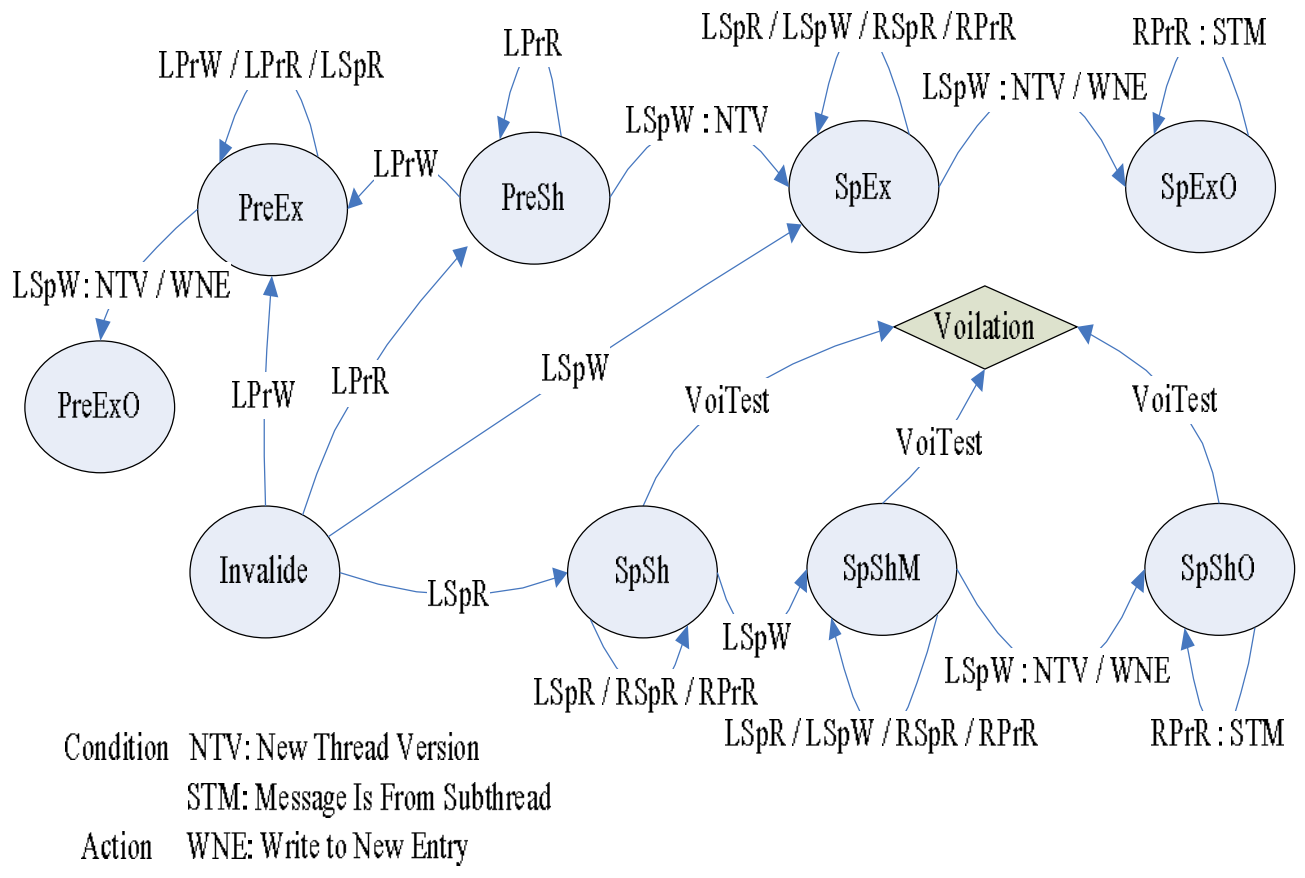

Fig.6 Memory Cache Accessing Mechanism

According to this mechanism we implement our memory cache line as shown in Fig.7. In our implementation, we separate the data with different version by thread version. A cache line with version 0 indicates it is generated in pre-computation. All the cache lines generated in speculation have a version greater than 0. We use old bit to denote whether the cache line is with the old version or not.

| Bit | Description |
|---|---|
| V | Valid |
| RL | Remote Loaded |
| M | Modified |
| Ver | Version |
| O | Old |

(a) cache line state bit

| State | V | RL | M | Ver | O |
|---|---|---|---|---|---|
| Invalid | 0 | X | X | X | X |
| PreSh | 1 | 1 | 0 | 0 | 1 |
| PreEx | 1 | 0 | 1 | 0 | 0 |
| PreExO | 1 | 0 | 1 | 0 | 1 |
| SpSh | 1 | 1 | 0 | TV | 0 |
| SpShM | 1 | 1 | 1 | TV | 0 |
| SpShO | 1 | 1 | 1 | TV | 1 |
| SpEx | 1 | 0 | 1 | TV | 0 |
| SpExO | 1 | 0 | 1 | TV | 1 |

(b) state encoding

| | | | | | | |
|---|---|---|---|---|---|---|
| V | RL | M | Ver | O | Address Tag | Data |

(c) cache line

Fig.7 Memory Cache Line Implementation

### 4.2.2 Register Cache Line

In Prophet, a thread can read or write a register in pre-computation or speculation. Table 5 enumerates all these operations. The basic states for a register cache are Validate (Va) and May-Commit (MC). By combination a register cache line in Prophet can be in one of the four states described in Table 6.

Table 5 Register Cache Operation

| operation | description |
|---|---|
| WP | Write in Pre-computation |
| RP | Read in Pre-computation |
| WS | Write in Speculation |
| RS | Read in Speculation |

Table 6 Register Cache State

| state | description |
|---|---|
| Init | the state when thread is spawned |
| MCommit | the value may be committed |
| Validate | the value should be validated |
| VaandMC | the value generated in pre-computation need validation and its newest version may be committed |

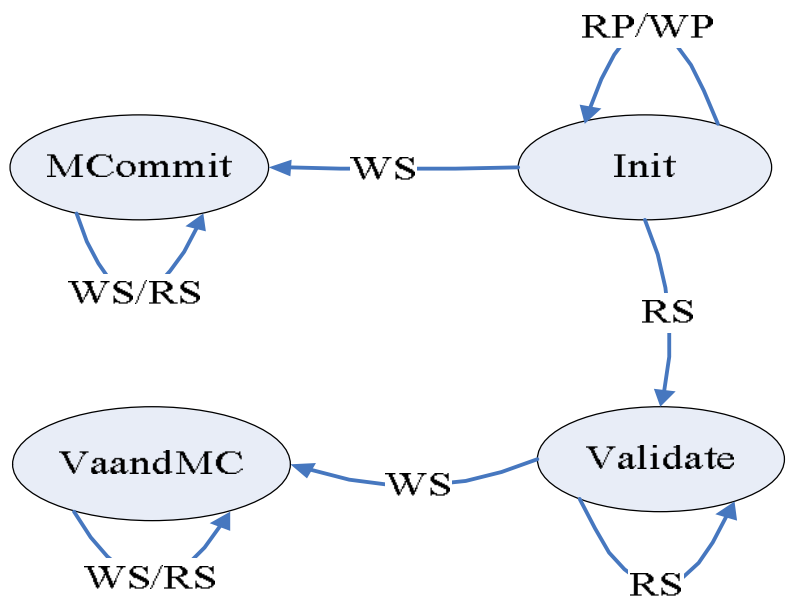

Fig.8 Register Cache Accessing Mechanism

| Bit | Description |
|---|---|
| V | Valid |
| L | Loaded First |
| M | Modified |

(a) cache line state bit

| State | V | L | M |
|---|---|---|---|
| Init | 1 | 0 | 0 |
| Validate | 1 | 1 | 0 |
| MCommit | 1 | 0 | 1 |
| VaandMC | 1 | 1 | 1 |
| InValid | 0 | X | X |

(b) state encoding

| | | | | |
|---|---|---|---|---|
| V | L | M | Tag | Data |

(c) cache line

Fig.9 Register Cache Line Implementation

We summarize our register cache mechanism in Fig.8. According to our scheme, when in pre-computation, the register is accessed directly without changing its state. When the thread quit pre-computation, all its data generated in pre-computation is saved in register cache and set up the Init state. When in speculation, if a register is read, its associated cache line state is transferred to Validate. If a register is written in speculation, its cache line state is transferred to May-Commit state, including MCommit state and VaandMC. In Prophet the register cache is implemented as shown in Fig.8.

### 4.3 Memory Data Dependence Violation Detection

Prophet capitalizes on a single snoopy bus protocol to detect memory data dependence violation in speculative execution (sp-execution) state. When a thread speculatively writes a data in sp-execution state, it will send a VioTest message to the bus. Its sub thread will pick up the message and parse the written address piggybacked along with the message to check whether a RAW violation has occurred in its own local L1 cache. If so, all the more speculative threads are squashed and the sub thread itself will be restarted.

Write operation in pre-computation state does not send a VioTest message, Prophet relies on verification to detect memory dependence violations that may occurred in pre-computation state. Before the stable thread commits, it will verify the speculative modifications generated by its immediate sub thread in pre-computation state. According to Fig.6 the cache lines needing verification will be in PreEx state or PreExO state. Verification is carried out just by comparing the values pre-computed by its sub thread with the ones generated by the stable thread. If all the values with the same address are equal, this means that no violation occurs. Otherwise all the sub threads are squashed and the stable thread will continue running.

### 4.4 Register Synchronization

Since p-slice can forecast the register values, Prophet doesn't transfer register values at runtime. The register transfer is postponed until thread committing. Just as the verification and committing of memory speculative modifications, register values are also verified by the stable thread and committed to its immediate sub thread. In Prophet, all the register values need verification are stored in the sub thread's register cache and are in Validate or VaandMC state. These values are read before they are written. If a register of the sub thread is in Init state which indicates it is neither read nor written, the stable thread will commit its value to the sub thread and the register value is synchronized.

## 5. Performance Evaluation

In order to evaluate our approach we develop the Prophet simulator which models a generic SpMT processor. Each processing element (PE) has its own program counter, fetch unit, decode unit, and execution unit that can fetch and execute instructions from a thread. In our evaluation, we assume that the architecture are fast enough so that thread spawning and squashing can be finished in 1 cycle. For benchmarks, we use 8 programs all written in C from Olden, and we add rook program which calculus the maximum number of chesses can put on a board with forbidden areas.

Table 7 presents static partition result generated by Prophet compiler for our benchmarks. Table 8, Table 9 and Table 10 shows the performance statistics under different amount of PEs separately. Fig.10 compares the speedup with different PE number. The average speedup achieved by Prophet is about 15.6% with 2 PEs, 34.5% with 4 PEs, and 43.3% with 8 PEs. Speedup will increase with the PE number moderately. By comparing in Fig.10, we can find that em3d can gain a dramatic upgrade in performance with increase of the PE number. This is because the data dependence violation can be effectively reduced by pre-computation and makes em3d always keep a high ratio of successful speculation. According to Table 7 we can also find that the programs with higher percentage of loop region apt to benefit from increase in PE number. Since the more loop region means the more parallelism can be exploited.

Table 7 Characteristics of the Olden benchmarks

| OLDEN | Thread Size | % of Dynamic Instr. of Diff. Thread Types | | p-slice Size | p-slice/Thread | Thread Live-ins |
|---|---|---|---|---|---|---|
| | | Loop | Non-loop | | | |
| mst | 35 | 55.5% | 44.5% | 3.8 | 10.8% | 2.8 |
| bh | 65.5 | 47.3% | 52.7% | 4.3 | 6.6% | 3.4 |
| power | 56.4 | 40.5% | 59.5% | 3.8 | 6.7% | 2.5 |
| tsp | 32 | 59.2% | 40.8% | 3.1 | 9.6% | 2.3 |
| em3d | 40.6 | 28.0% | 72% | 4.9 | 12.1% | 3.8 |
| bisort | 35.4 | 73.7% | 26.3% | 3.5 | 9.9% | 2.2 |
| voronoi | 84.4 | 55.8% | 44.2% | 4.2 | 6% | 3.2 |
| rook | 32 | 50% | 50% | 3.3 | 10.3% | 2.3 |
| health | 38.8 | 22.2% | 77.8% | 3.4 | 8.8% | 2.4 |

Table 8 Performance Statistic (with 2 PEs)

| OLDEN | Spawned Threads | Failed Threads | % of Successful Threads | Speedup |
|---|---|---|---|---|
| mst | 305 | 213 | 30% | 2.8% |
| bh | 3811 | 1526 | 60% | 3% |
| power | 13234 | 5785 | 56% | 4.5% |
| tsp | 14 | 2 | 86% | 9% |
| em3d | 2287 | 266 | 88% | 52% |
| bisort | 137156 | 104388 | 24% | 3% |
| voronoi | 754 | 631 | 16.3% | 15.5% |
| rook | 655 | 405 | 38% | 23% |
| health | 2979 | 1256 | 58% | 28% |

Table 9 Performance Statistic (with 4 PEs)

| OLDEN | Spawned Threads | Failed Threads | % of Successful Threads | Speedup |
|---|---|---|---|---|
| mst | 666 | 523 | 21.5% | 4.4% |
| bh | 26885 | 12482 | 53.6% | 12.8% |
| power | 42961 | 14520 | 66.0% | 14.3% |
| tsp | 29 | 14 | 52% | 17.3% |
| em3d | 4297 | 756 | 82.4% | 134.2% |
| bisort | 251572 | 208564 | 17% | 9% |
| voronoi | 1664 | 1540 | 7.5% | 15.8% |
| rook | 1185 | 802 | 32.3% | 47.5% |
| health | 6182 | 3527 | 43% | 50.3% |

Table 10 Performance Statistic (with 8 PEs)

| OLDEN | Spawned Threads | Failed Threads | % of Successful Threads | Speedup |
|---|---|---|---|---|
| mst | 1172 | 1029 | 12.2% | 4.3% |
| bh | 34213 | 21897 | 36.0% | 15.5% |
| power | 46434 | 17993 | 61.3% | 14.3% |
| tsp | 39 | 24 | 38% | 17.3% |
| em3d | 5873 | 2132 | 64% | 211% |
| bisort | 392861 | 349853 | 11% | 9% |
| voronoi | 3363 | 3239 | 3.9% | 16.4% |
| rook | 1755 | 1341 | 23.6% | 50.6% |
| health | 8052 | 5397 | 33% | 50.9% |

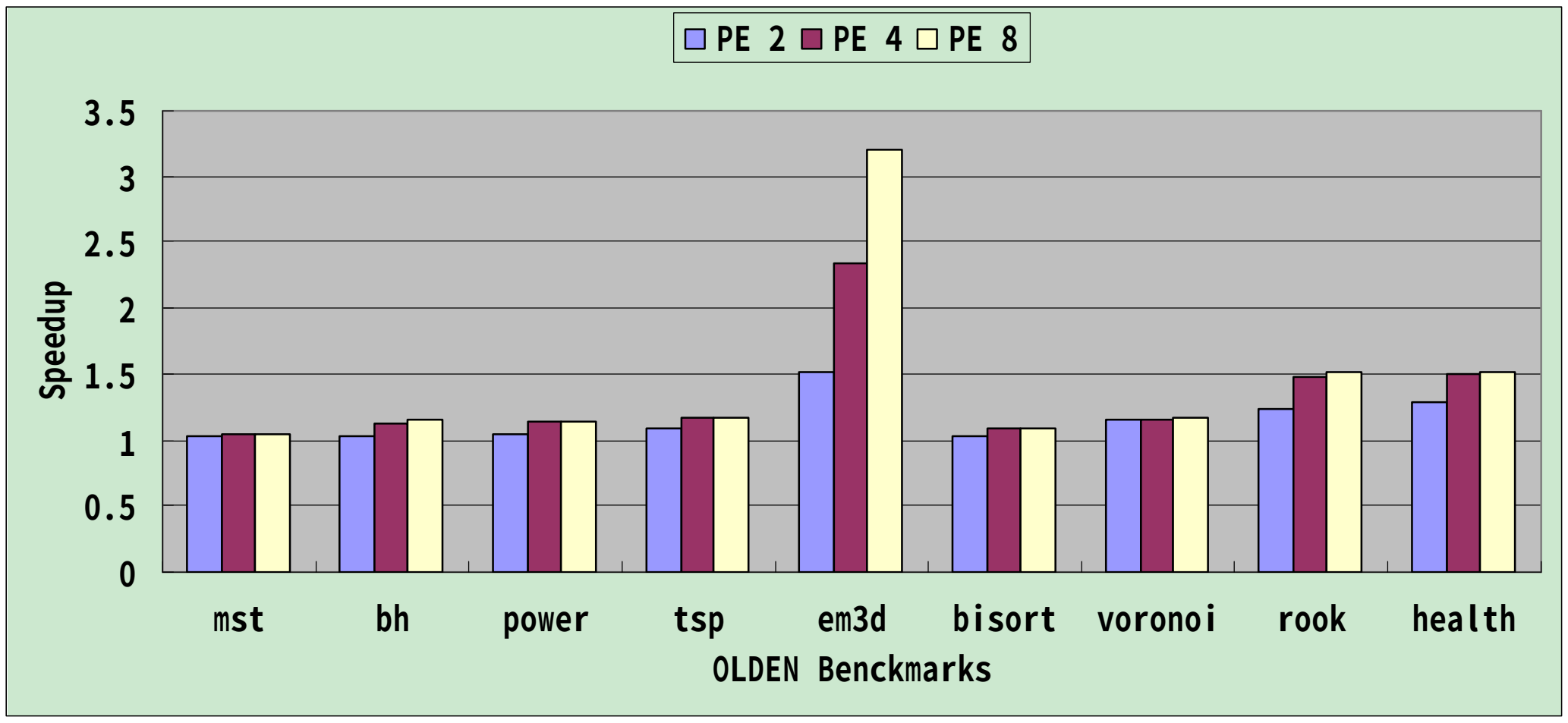

Fig.10 Speedup comparison under different PE number

## 6. Summary

We have presented the Prophet speculative multi-threading execution model with architectural support based on CMP. It focuses on exploiting thread level parallelism and uses pre-computation to reduce inter-thread data dependence violations. Its key features are as follows.

1) Prophet execution model uses pre-computation to reduce data dependence violation. Each thread is associated with a pre-computation slice (p-slice) to predict its live-ins.
2) The register cache is designed to buffer the live-in data

for registers. Register value is validated and committed to its sub thread so there is no runtime cost for register synchronization.

3) A multi-version cache is designed to buffer speculative data with multi-version not only among different threads but also within each individual thread.

Prophet is a speculative multithreading execution model depending on software and hardware co-design. Our simulation-based evaluation has validated its correctness and shows that with 4 cores Prophet execution model can outperform the program's sequential execution by 34.5% on average. The speedup can increase with the number of cores in the system. But the range of increase relies on the program's intrinsic attributes. If the RAW dependency violations can be eliminated by pre-computation, the speedup can keep raising up in a considerable range while increasing the core number.

Our current work focuses on the optimization of architecture support, attempting to reduce the overhead of spawning, validation, commitment and restart of speculative threads at run-time. A FPGA implementation of Prophet architecture is also under going. And a relative goal is the system level evaluation.